\documentclass[11pt]{article} 
\usepackage{amssymb}
\usepackage{amsmath}
\usepackage{latexsym}
\usepackage[mathscr]{eucal}
\usepackage{citesort}
\usepackage{url}

\newlength{\tagwidth}
\usepackage{a4}
\def\R{\mathbb R}

\def\U{\mathbb U}

\newcounter{mnotecount}[section]

\renewcommand{\themnotecount}{\thesection.\arabic{mnotecount}}

\newcommand{\mnote}[1]
{\protect{\stepcounter{mnotecount}}$^{\mbox{\footnotesize
$
\bullet$\themnotecount}}$ \marginpar{
\raggedright\tiny\em $\!\!\!\!\!\!\,\bullet$\themnotecount: #1} }
\begin{document}
{BOUNDARY CONDITIONS AT SPATIAL INFINITY

FROM A HAMILTONIAN POINT OF VIEW\footnote{Published in {\em
``Topological Properties and Global Structure of
      Space-Time"}, ed. by P. Bergmann, V. de Sabbata, pp.
      49-59, Plenum Press, New York 1986 }}

\bigskip

Piotr T.~CHRU\'SCIEL

Institute for Theoretical Physics

Polish Academy of Sciences

Warsaw, Poland

\bigskip

\parindent=0 mm
INTRODUCTION

\bigskip

There are many, both conceptually and technically different ways
to obtain the ADM expression for the energy of the gravitational
field, some of the published methods containing inconsistencies,
most of them raising doubts about uniqueness of the final result.
The author wishes to present here a simple way of obtaining this
expression in a geometrical Hamiltonian setting allowing for an
exact analysis of all ambiguities present. One of the results of
this study is a considerable weakening of the boundary conditions
at spatial infinity, for which the energy-momentum of an initial
data set is finite and well defined. The derivation of the ADM
Hamiltonian presented here is the simplest one known to the
author, as far as calculations are concerned.

The starting point of our derivation of the ADM Hamiltonian will
be the so-called affine formulation of general relativity [1].
Many different formulations of general relativity may be used for
this purpose, this is however in the general affine framework that
the calculations are the simplest ones. It must be noted that
vacuum general relativity is a somewhat pathological theory in
this context (cf. the discussion following formula (1)), but this
presents no difficulty in our approach: we shall start with a
theory which contains a certain number of matter fields, find the
Hamiltonian for this theory, and finally set the non-gravitational
fields to zero, obtaining thus the Hamiltonian for vacuum general
relativity. All the calculations required to obtain the final
expression for the Hamiltonian are four-dimensional covariant,
they do not necessitate a 3 + 1 decomposition of the fields. This
requires some justification, because usually the phase space of
general relativity is thought of as the space of functions
$(g_{ij}, P^{ij})$ (the ADM data) on a three dimensional manifold,
satisfying certain constraint equations and certain boundary
conditions. However, by well known evolution theorems, every such
set of functions gives rise to a four-dimensional Lorentzian
manifold in which the four-dimensional field equations are
satisfied  -  this shows that the phase space of general
relativity (or, in fact, of any field theory)  is isomorphic to
the space of solutions of the equations of the theory, satisfying
certain boundary conditions. In general relativity the ADM data
give one possible parameterisation of this space (which is
incomplete, since constraint equations are still present). Anybody
acquainted with symplectic geometry knows that no coordinates are
required to make sense of Hamilton's equations of motion (or,
equivalently, any set of coordinates is fine) once the symplectic
structure on the phase space is given. This leads one to expect
that there should exist a framework in which four dimensional
covariant quantities can be used in the calculations - such a
framework has been recently constructed by J. Kijowski and W.
Tulczyjew [2]. We shall not attempt here to review this
construction and will just present how it works in general
relativity.

\bigskip
\parindent=0 mm
THE AFFINE FORMULATION OF GENERAL RELATIVITY

\bigskip
\parindent=12 mm
As has been shown by M. Ferraris and J. Kijowski [1,3,4], every
Lagrangian theory of gravitation and some matter fields $\phi^A$
satisfying field equations deriving from an action of the form
$$
I [g_{\mu\nu},\phi^A]=\int (g^{\mu\nu} R_{\mu\nu}+L_m(\phi^A,
\phi^A{}_{,\mu}, g_{\mu\nu}, g_{\mu\nu,\sigma}))(-det\ g)^{1/2}
d{}^4 x\ \eqno(1)
$$
\parindent=12 mm
can be formulated as a ``purely affine theory" in the following
sense: the theory may be considered as a theory of a $GL (4,R)$
connection field $\Gamma_{\mu\nu}^\lambda$, the field equations
deriving from an action of the form: $$I
[\Gamma_{\mu\nu}^\lambda,\phi^A]=\int L(\Gamma_{\mu\nu}^\lambda,
\Gamma_{\mu\nu,\sigma}^\lambda,\phi^A,\phi^A{}_{,\sigma})\ d^4
x.$$
 (Care must be taken when interpreting this result. The scalar
density $L$ appearing above is what the physicists call a
Lagrangian only in the case of no constraints in the
 ``infinitesimal configuration space", the reader is referred to
ref.~[2] for details. In the case of vacuum general relativity $L$
above is defined only on the constraint hypersurface
$R_{\mu\nu}=0$, its numerical value being zero. Let us also note,
that Kijowski's theorem does not hold in presence of fermionic
fields, because there is no natural formulation of such theories
with an action of the form (1)).

\bigskip
\parindent=12 mm
It has also recently been shown [5], that every purely affine
theory with a Lagrangian of the form
$$L=L(R^\lambda{}_{\mu\nu\rho})\ \eqno(2)$$
\parindent=0 mm
can be interpreted as an Einstein theory of gravitation, in which
certain components of the connection can be considered as tensor
matter fields. For example, in the simple case of a Lagrangian
taking the form
$$L=L(K_{\mu\nu},F_{\mu\nu}),$$
where

$$K_{\mu\nu}=(R^\alpha{}_{\mu\alpha\nu}+R^\alpha{}_{\nu\alpha\mu})/2\;,\qquad F_{\mu\nu}
=R^\alpha{}_{\alpha\mu\nu}\;.$$

the metric is obtained from the equation
$$((-det \ g)^{1/2}  g^{\mu\nu})/16\pi = \pi^{\mu \nu},\ \eqno(3)$$
$$
\pi^{\mu \nu}=\partial L/\partial K_{\mu \nu}.\ \eqno(4)
$$
When $L$ is taken to be of the form $$L= (-det \ K_{\alpha
\beta})^{1/2} \  K^{\mu\alpha} \ K^{\nu\beta} \ F_{\mu\beta} \
F_{\nu\alpha} / 16\pi,$$
 where $K^{\alpha\beta}$ is the inverse
tensor to $K_{\alpha\beta}$, the field equations are just
Einstein-Maxwell equations [6,7] (in this case the quantity
$A_\mu=\Gamma_{\mu\lambda}^\lambda$ has the interpretation of the
electromagnetic potential). Kijowski and Tulczyjew [2] have
derived the following formula for the Hamiltonian of the theory:
$$
E(X,\Sigma) = \int_{\Sigma} (\pi_\alpha{}^{\gamma\mu\beta} {\cal
L}_X \Gamma_{\beta\gamma}^\alpha - X^\mu L) \eta_\mu,\ \eqno(5)
$$
where $\Sigma$ is any hypersurface of codimension 1 in the
manifold $M$ on which we study the dynamics of the gravitational
field, $X$ is any vector field on $M$, and
$$
\pi_\lambda{}^{\mu\nu\rho}=\partial L/\partial
\Gamma_{\rho\mu,\nu}^\lambda \ \eqno(6)
$$
(the ``strange-looking" positioning of indices on
$\pi_\alpha{}^{\beta\gamma\delta}$ in (5) and (6) comes from the
conventions of ref.~[8], which are used throughout this paper). We
will show that $E$ is indeed a Hamiltonian for translations
generated by $X$, modulo some boundary terms which will be
analysed later on. We will restrict ourselves to Lagrangians of
the form (2). As has been pointed out above, this form of $L$ is
general enough to include the Einstein-Maxwell theory, and
therefore sufficient to obtain what we finally aim to: the
Hamiltonian for vacuum general relativity. To show that formula
(4) provides a Hamiltonian on the phase space (it is, the space of
fields satisfying the field equations, and some boundary
conditions to be imposed later on) let us calculate the
differential of $E$:
\begin{eqnarray*}
\delta E(X,\Sigma)&=& \int_{\Sigma}(\
\delta\pi_\alpha{}^{\gamma\mu\beta}{\cal L}_X
\Gamma_{\beta\gamma}^\alpha  +
\pi_\alpha{}^{\gamma\mu\beta}\delta{\cal L}_X
\Gamma_{\beta\gamma}^\alpha - X^\mu \delta L\ ) \eta_\mu \\ &=&
\int_{\Sigma}(\ {\cal L}_X  \Gamma_{\beta\gamma}^\alpha \delta
\pi_\alpha{}^{\gamma\mu\beta}  - {\cal L}_X
\pi_\alpha{}^{\gamma\mu\beta} \delta\Gamma_{\beta\gamma}^\alpha \
) \eta_\mu \\&&+ \int_{\Sigma}\left(\ {\cal L}_X
\pi_\alpha{}^{\gamma\mu\beta}\delta\Gamma^\alpha_{\beta\gamma}
  - X^\mu (\pi_\alpha{}^{\gamma\sigma\beta}\delta\Gamma_{\beta\gamma}^\alpha)_{,\sigma}\right)\eta_\mu
,\ \qquad\mbox{ \hfill\rm (7)}
\end{eqnarray*}
 and we have used the formula
$$\delta L = (\pi_\alpha{}^{\gamma \sigma \beta}
\delta\Gamma_{\beta\gamma}^\alpha)_{,\sigma}\;,$$
 which holds in virtue of field
equations. It is easily seen (using the definition of Lie
derivatives) that the last integral in the right hand side of
formula (7) is a total divergence, and one obtains
$$
\delta E = \int_{\Sigma} (\ {\cal L}_X \Gamma_{\beta\gamma}^\alpha
\delta\pi_\alpha{}^{\gamma\mu\beta}-{\cal L}_X
\pi_\alpha{}^{\gamma\mu\beta} \delta\Gamma_{\beta\gamma}^\alpha\
)\ \eta_\mu + \int_{\partial\Sigma} \pi_\alpha{}^{\gamma\beta[\mu}
X^{\nu]} \delta\Gamma_{\beta\gamma}^\alpha \eta_{\mu\nu}.\
\eqno(8)
$$
This formula has a deep symplectic meaning, for details the reader
is referred to ref.~[2]. It can be used as a starting point of the
canonical analysis of general relativity [9,10,11] and in fact it
``looks like" Hamilton's equations of motion
$$dH = \dot q\ dp-\dot p\ dq\;,$$
apart from the boundary term.
\bigskip
\parindent=12 mm
The numerical value of $E$ is given by equation (5) - it takes a
three lines calculation to show that $E$ is a boundary integral,
and to calculate its actual value [8]:
$$
E=\bigl(\int_{\partial\Sigma} \pi_\delta{}^{\mu\alpha\beta}
(X^\delta{}_{;\mu} + (\Gamma_{\sigma\mu}^\delta
-\Gamma_{\mu\sigma}^\delta) X^\sigma) \eta_{\beta\alpha}\bigr)/2.\
\eqno(9)$$ In what follows we shall confine our attention to pure
gravity, in which case $\pi_\lambda{}^{\mu\nu\alpha}$ takes the
form [8]
$$
\pi_\lambda{}^{\mu\nu\alpha}=2\
\pi^{\mu[\alpha}\delta_\lambda^{\nu]}, \ \eqno(10)
$$
$\pi^{\mu\nu}$ is related to the metric via (3) and, as a
consequence of the field equations, $\Gamma^\lambda_{\mu\nu}$ is
the symmetric Riemannian connection of $g_{\mu\nu}$. It is
convenient to introduce the variable
$$A_{\mu\nu}^\lambda=\Gamma_{\mu\nu}^\lambda-\delta_\mu^\lambda\Gamma_\sigma^\sigma.$$
Insertion of (10) into (8) leads to
$$\delta E = \theta_\Sigma + \int_{\partial \Sigma}
\pi^{\alpha\beta} X^{[\mu} \delta A^{\nu]}_{\alpha\beta}
\eta_{\mu\nu}\;,$$
$$\theta_\Sigma= \int_\Sigma({\cal L}_X A^\mu_{\alpha\beta} \delta
\pi^{\alpha\beta} - {\cal L}_X\pi^{\alpha\beta} \delta
A^\mu_{\alpha\beta})\eta_\mu\;. \ \eqno(11)$$ In the pure vacuum
case we are considering here, insertion of (10) into (9) leads to
the Komar integral:
$$E=\bigl(\int_{\partial\Sigma}\nabla^\mu X^\nu \eta
_{\mu\nu}\bigr)/16\ \pi$$
(it is worthwhile noting, that this is the Komar expression with a
 ``factor wrong by 1/2" - when evaluated for the Schwarzschild
metric with $X=\partial/\partial t$, the Killing vector, it gives
m/2).

\bigskip
\parindent=0 mm
ASYMPTOTICALLY FLAT SPACE-TIMES

\bigskip
\parindent=12 mm
In order to analyse the boundary terms in eq.\ (11) let us assume
that $\Sigma$ is a spacelike hypersurface extending up to infinity
in an asymptotically flat space-time, where ``asymptotic flatness"
is to be understood as follows: outside a world tube there exists
a coordinate system such that
$$
g_{\mu\nu}=\mathring{\eta}_{\mu\nu}+h_{\mu\nu}, \ \eqno(12)
$$
where $\mathring{\eta}_{\mu\nu}$ is the Minkowski metric, and
$h_{\mu\nu}$ satisfies
$$
|h_{\mu\nu}| \leq  C/r^\alpha\;,\qquad |h_{\mu\nu,\sigma}| \leq
C/r^{\alpha+1}, \ \eqno(13)
$$
for some $\alpha$ to be specified later. It will be assumed, that
$X$ tends asymptotically to the vector $\partial/\partial  t$. If
we want the functional E to generate the time translations (it is,
translations along $X$) we have to ``kill" the boundary
non-dynamical terms in formula (11) - and this can be done by
imposing appropriate boundary conditions in the space of metrics
we are working with. Formula (13) shows, that
$$\pi^{\alpha\beta}X^{[\mu} \delta A^{\nu]}_{\alpha\beta}\sim
1/r^{\alpha+1}.$$ If we required the boundary terms to vanish in
the limit $r \rightarrow  \infty$, we should have $\alpha > 1$ but
then, due to the positive energy theorems, the metric $g_{\mu\nu}$
would have to be flat. This is related to the fact, that $E$ is a
Hamiltonian for time-translations in a space of functions, where
certain leading order components of $\Gamma_{\mu\nu}^\lambda$ are
kept fixed on the boundary  --- but this is not the way we have
introduced asymptotically flat metrics. In (12) we are keeping
fixed the leading order components of the metric, and not of the
connection. A remedy to this problem is given by the following
procedure: introduce, for large $r$, a fixed ``background" metric
$f_{\mu\nu}$ and let
$$\mathring{A}_{\mu\nu}^\alpha=\mathring{\Gamma}_{\mu\nu}^\alpha-\delta_\mu^\alpha
\mathring{\Gamma}_{\nu\sigma}^\sigma\;,$$ where
$\mathring\Gamma_{\beta\alpha}^\alpha$ are the Christoffel symbols
of the metric $f_{\mu\nu}$. Introduce
$$
H = E - \int_{\partial\ \Sigma} \pi^{\alpha\beta} X^{[\mu}
D_{\alpha\beta}^{\nu]} \ \eta_{\mu\nu}, \ \eqno(14)
$$
where
$$D_{\beta\gamma}^\alpha = A_{\beta\gamma}^\alpha -
\mathring A_{\beta\gamma}^\alpha$$ $(D_{\beta\gamma}^\alpha$ is a
tensor). From (11) and (14) one finds
$$\delta H = \delta E - \int_{\partial\Sigma} \delta \pi^{\alpha \rho}
X^\mu D_{\alpha\beta}^\nu \ \eta_{\mu\nu}-\int_{\partial\Sigma}
\pi^{\alpha \beta}X^\mu \delta D_{\alpha\beta}^\nu \
\eta_{\mu\nu}\;.$$ Since the background is fixed, $\delta
D_{\beta\gamma}^\alpha=\delta A_{\beta\gamma}^\alpha$, therefore
$$
\delta H = \theta_\Sigma + \int_{\partial\Sigma}\ X^\mu
D_{\alpha\beta}^\nu \delta \pi^{\alpha\beta}\ \eta_{\mu\nu}.\
\eqno(15)
$$
If we consider metrics satisfying (13) we have $$
|\delta\pi^{\alpha\beta}|\le C r^{-\alpha}\;, \quad
|D_{\beta\gamma}^\alpha| \le C r^{-\alpha-1}\;,$$
$$
|X^\mu D_{\alpha\beta}^\nu \delta \pi^{\alpha\beta}| \le C
r^{-\alpha-1}\;.
$$ The non-dynamical terms in (15) will give no contribution if we
require $2 \alpha + 1 > 2$, it is $\alpha = 1/2 + \varepsilon,
\varepsilon$ being any strictly positive number. In this space of
metrics we will simply have
$$\delta H = \theta_\Sigma \;.$$
The final formula for the Hamiltonian can be written in the form
[12]:
$$
H = \bigl(\int_{\partial \Sigma}\
E^{\alpha\beta}\eta_{\alpha\beta}\bigr)/\ \ 16 \pi,\ \eqno(16)
$$
where \begin{eqnarray*} E^{\alpha\beta}&=& (U_\mu{}^{\alpha\beta}
X^\mu + g^{\lambda[\alpha}\delta^{\beta]}_\mu X^\mu{}_{|\lambda})
(-\det g)^{1/2}
\\U_\nu{}^{\alpha\beta}&=&g_{\nu\mu}(e^2
g^{\mu[\alpha}g^{\beta]\sigma})_{|\sigma}e^{-2}\;,\qquad e^2 =
\det g_{\mu\nu}/ \det f_{\mu\nu}\qquad\qquad \mbox{\rm(17)}
\end{eqnarray*}
and a
bar refers to covariant differentiation with respect to the
background metric\footnote{ The introduction of the background
metric is motivated by the way we defined asymptotic flatness, and
also by the fact that we want the integrand of H to have correct
transformation properties. Instead of considering (14) we could
consider
$$H' = E - \int_{\partial \sum} \pi^{\alpha\beta} X^{\mu}
A^{\nu}_{\alpha\beta} \eta_{\mu\nu}$$ Although the integrand of
$H'$ ceases to be a two-form density from a four-dimensional point
of view, it can be shown that it is intrinsically defined by $X$
and the geometry of $\partial \Sigma$. The ``background metric"
approach seems however more convenient for further purposes.}. It
may be of some interest to mention, that the transition from $H'$
to $H$ is accomplished via a sort of Legendre transformation. A
good analogy is provided by thermodynamics, where one defines the
internal energy $$dU = TdS + pdV$$ and one interprets the
increments of $U$ as the amount of energy required to change the
state of, say, a gas by heating it while keeping its volume fixed.
Another type of energy (the enthalpy) is obtained if one heats the
gas while keeping its pressure fixed:
$$dH = d(U-pV) = TdS - Vdp\;.$$

Let us briefly analyse the expression (17). It must be emphasised,
that it provides a Hamiltonian for the dynamics of the
gravitational field for any background, provided the boundary
integral in (15) vanishes - in particular, one can use it for
asymptotically anti-de-Sitter space-times. It is also worthwhile
noting, that the vector field $X$ is still arbitrary in this
formula. It seems that the discussion of the dependence of $ H$
upon $X$ and the background metric has to be done separately for
each class of space-times considered --- from now on we will
restrict the discussion to the usual dynamical description of
asymptotically flat space-times, in the sense of (13), with
$\alpha > 1/2$. Let us therefore specify $\Sigma$ to be a $t =$
const hypersurface, $f_{\mu\nu}$ to be the flat metric
$\mathring{\eta}_{\mu\nu}$, and $X$ to be any translational
Killing vector of the metric $\mathring{\eta}_{\mu\nu}$. Since $X$
is now ``background covariantly constant" the second term in (17)
vanishes, therefore
$$H(X) =\bigl(\int_{\partial \Sigma}(-\det \ g)^{1/2}\ U_{\mu}{}^{\alpha\beta}
\ X^\mu \ \eta_{\alpha\beta})/16\ \pi\;.$$ This formula is known
as the ``Freud superpotential" for the ``Einstein energy-momentum
pseudo-tensor". In the ADM notation it takes the following form
(in the coordinates $x^\mu$ satisfying (13), with $\alpha > 1/2$):
$$
P_0 = H(X=\partial/\partial t) = \bigl(\int_{\partial\Sigma}
(g_{ik,k}-g_{kk,i}) \ dS_i\ \bigr)/ 16\ \pi, \ \eqno(18)
$$
$$
P_i = H(X=\partial/\partial x^i) = \bigl(\int_{\partial\Sigma} P^{ij}\ dS_j\ \bigr)/ 8\ \pi. \ \eqno(19)
$$
The above formulae are the well known ADM expressions for the
energy-momentum of an initial data set.

While inspecting formulae (18) and (19) three questions arise
immediately:

1) what does the symbol $\int_{\partial\Sigma}$ mean ? Such an
integral is usually understood as the limit of integrals over
spheres, while the radii of spheres tend to infinity. Does such a
limit exist and, if so, does it depend upon the family of spheres
used to perform this calculation?

2) If these limits exist in some sense, are they finite ?

3) Do these limits depend upon the particular system of
coordinates (satisfying (13)) used to perform the calculations ?
Neither (18) nor (19) are defined in an intrinsic way on $\Sigma$
--- (19) contains a free vector index, and (18) contains partial
derivatives of a tensor.

We will analyse these problems in the case of a fixed Cauchy
hypersurface $\Sigma$ (it can be shown, that $P_\mu$ transforms as
a Lorentz convector under boosts of $\Sigma$, this will however be
discussed elsewhere). Before giving the precise statement of the
theorems, it is useful to introduce first some terminology.
Suppose one is given a pair $(g,\phi)$, where

1) $g$ is a Riemannian metric on a three dimensional manifold $N$,
$N$ diffeomorphic to $\R^3\setminus B(R)$, where $B(R)$ is a
closed ball ($N$ can be thought of as one of (possible many)
``ends" of $\Sigma$\ ).

2) $\phi$ is a coordinate system in the complement of a compact
set $K$ of $N$ such that, in local coordinates $\phi^i (p)=x^i$
the metric takes the following form:
$$
g_{ij}=\delta_{ij}+k_{ij}, \ \eqno(20)
$$
and $k_{ij}$ satisfies
$$
\forall_{i,j,k,x}\ |k_{ij} (x)| \leq \ C/(r+1)^{\alpha}\ \, \ \
|\partial k_{ij}/\partial x^k(x)| \leq \ C/(r+1)^{\alpha+1}\ \, \
\ (\ r(x) = (\sum (x^i)^2\ )^{1/2}\ ), \ \eqno(21)
$$
for some constant $C \in R$. Such a pair $(g,\phi)$ will be called
$\alpha$--admissible. Let us restate the remaining boundary
conditions (13) in the ADM language:
$$\forall i,j,x \quad|(N-1)(x)| \le C/(r+1)^\alpha\;,\qquad
|N^i(x)|\le C/(r+1)^\alpha\;,$$
$$ |N_{,i}(x)| \le C/(r+1)^{\alpha+1}\;,\quad
|P_{ij}(x)|\le C/(r+1)^{\alpha+1}\;,\quad |N^i{}_{,j}(x)|\le
C/(r+1)^{\alpha+1}\;. \eqno(22)$$

\noindent\underbar{Theorem 1}: Suppose that

1) $(g,\phi)$ is $\alpha$--admissible, with $\alpha > 1/2$,

2) the conditions (22) are satisfied,

3) $(g_{ij}, P_{ij})$ satisfy the constraint equations, with
integrable sources.

Let $S(R)$ be any one-parameter family of differentiable spheres,
such that $r(S(R))=\min _{x\in S(R)} r (x)$ tends to infinity, as
$R$ does. Define \begin{eqnarray*}
 m(g,\phi) & = & \lim_{R\to\infty} \frac 1 {16\pi} \int_{S(R)}
 (g_{ik,i}-g_{ii,k})dS_k\;,
 \\
 P_i(g,\phi) & = & \lim_{R\to\infty} \frac 1 {8\pi} \int_{S(R)}
 P^{ij}dS_j
 \end{eqnarray*}
  (these
integrals have to be calculated in the local $\alpha$--admissible
coordinates $\phi^i (p)=x^i$). $m$ and $P_i$ are finite,
independent upon the particular family of spheres $S(R)$ chosen,
provided $r(S(R))$ tends to infinity as $R$ does.

\bigskip

\noindent\underbar{Proof}: Let $f_{\mu\nu}$ be the flat metric
$ds^2=-dt^2+\Sigma(dx^i)^2$ (where the $dx^i$ refer to the local
coordinate system $\phi^i$ on $\Sigma$ ). Define
$$A(X,R,f)=\int_{S(R)} U_\mu{}^{\alpha\beta} X^\mu\ \ (-\det g)^{1/2}
\eta_{\alpha\beta}\;.$$
 The Einstein-von Freud identity takes the
following form:
\begin{eqnarray*}
A(X,R_2,f) - A(X,R_1,f) &= & \int_{\Gamma(R_1,R_2)} \mbox{
(``expression quadratic in $(\Gamma_{\mu\nu}^\lambda -
\Gamma_{\mu\nu}^\lambda)$"} \\ && + \mbox{``expression linear in
$T_{\mu\nu}$")}\; d^3 x\;.\end{eqnarray*} (see, for example,
ref.~[12] for the explicit form of the volume integrand), where
$\Gamma (R_1,R_2)$ is the ``annulus" lying between $S(R_1)$ and
$S(R_2)$. For $X=\partial/\partial t$ or $X=\partial/\partial
x^i$, and $r(S(R_2)) > r(S(R_1))$ the volume integral above is
bounded by a constant independent of $R_2$ in virtue of our
hypotheses, tending to zero as $R_1$ tends to infinity. The reader
may easily establish all the claimed properties of $m(g,\phi)$ and
$P_i(g,\phi)$ using this observation.

\bigskip

In the proof of the theorem to follow we will need the following
simple lemma:

\bigskip
\parindent=0 mm
\underbar{Lemma 1}: Let $(g,\phi_1)$ and $(g,\phi_2)$ be
$\alpha_1$ and $\alpha_2$--admissible, respectively, with any
$\alpha_a
> 0$. Let $\phi_1 \circ\phi_2^{-1}: \R ^3\backslash K_2 \rightarrow \R ^3\backslash K_1$
be a twice differentiable diffeomorphism, for some compact sets
$K_1$ and $K_2 \subset \R^3$. Then, in local coordinates
$$\phi^i_1 (p)=x^i\,\qquad \phi_2^i (p)=y^i\;,$$
the diffeomorphisms $\phi_1 \circ \phi_2^{-1}$ and $\phi_2 \circ
\phi_1^{-1}$ take the form
$$x^i(y) = \omega^i{}_j\ y^i+\eta^i (y)\;,\qquad  y^i (x) = (\omega^{-1})^i{}_j\
x^i\ + \zeta^i (x)\;,$$ $\zeta^i$ and $\eta^i$ satisfy, for some
constant $C \in \R$,
\begin{eqnarray*}
&&|\zeta^i(x)| \le C (r(x)+1)^{1-\alpha}\;, \qquad
|\zeta^i{}_{,j}(x)| \le C (r(x)+1)^{-\alpha}\;,
\\ &&|\eta^i(y)| \le C (r(y)+1)^{1-\alpha}\;, \qquad |\eta^i{}_{,j}(y)|
\le C (r(y)+1)^{-\alpha}\;,
\\ && r(x) = (\sum (x^i)^2)^{1/2}\;,\qquad  r(y) = (\sum (y^i)^2)^{1/2}\;,
\end{eqnarray*}
 with $\alpha=\min (\alpha_1, \alpha_2), \omega^i{}_j$ is an $O(3)$
matrix, and $r^0$ is to be understood as $\ln r$.

\bigskip

\parindent=0 mm
\underbar{Proof}: This lemma is intuitively obvious, there are
however a few technicalities needed to make the proof
mathematically rigorous. Let us first note, that both $(g,\phi_1)$
and $(g,\phi_2) $ are $\alpha$--admissible, so that we do not have
to worry about two constants $\alpha_1$ and $\alpha_2$. In (21) we
can also take a common constant $C =\max (C_1, C_2)$. Let
$g_{ij}^1$ and $g_{ij}^2$ be the representatives of g in local
coordinates $\phi_1$ and $\phi_2$. (21) implies, that $g_{ij}^1$
and $g_{ij}^2$ are ``uniformly elliptic", it is there exist
positive constants $C'_1$ and $C'_2$ such that
$$
\forall {X^i} \in \R^3\quad \forall x \in \R^3 \setminus K_a\qquad
(C'_a){}^{-1} \Sigma (X^i)^2 \leq g_{ij}^a\ X^i X^j \leq C'_a \Sigma
(X^i)^2\, \ \eqno(24)
$$
$a=1,2$. From now on $C,C'$, etc. will denote constants which may
vary from line to line, their exact values can be estimated at
each step but are irrelevant for further purposes. Let us write
down the equations following from the transformation properties of
the metric
$$
g_{ij}^2 (y)=g_{k\ell}^1 (x(y)) {\partial x^k\over \partial y^i} \
{\partial x^\ell\over \partial y^j}\, \ \eqno(25)
$$
$$
g_{ij}^1 (x)=g_{k\ell}^2 (y(x)) {\partial y^k\over \partial x^i} \
{\partial y^\ell\over \partial x^j}\;. \ \eqno(26)
$$
Contracting (25) with $g_1^{ij}$ ((26) with $g_2^{ij}$) and using
the uniform ellipticity of $g_{ij}^1$ ($ g_{ij}^2$) one obtains
$$
\displaystyle\sum_{k,i} \bigl|{\partial x^k\over \partial
y^i}\bigr| \leq C \;,\qquad \displaystyle\sum_{k,i}
\bigl|{\partial y^k\over \partial x^i}\bigr| \leq C \;. \
\eqno(27)
$$
Inequalities (27) show that all the derivatives of $x(y)$ and
$y(x)$ are uniformly bounded. Let $\Gamma_x$ be the ray joining
$x$ and $K_1$, and let $y_0^i (x)$ be the image by $\phi_2 \circ
\phi_1^{-1}$ of the intersection point of $K_1$ with $\Gamma_x$
(if there is more than one, choose the one which is closest to
$x$). We have, in virtue of (27)
$$|y^i(x)-y_0^i(x)| = |\int_{\Gamma_x} (\partial y^i / \partial
x^k) dx^k| \leq C\ r(x)\;,$$
so that
$$
r(y(x)) \leq C\ r(x) + C^-\;. \ \eqno(28)
$$
A similar reasoning shows
$$
r(x(y)) \leq C\ r(y) + C^-\;. \ \eqno(29)
$$
(28) and (29) can be combined into a single inequality
$$
r(y(x))/C - C^- \leq r(x) \leq C\ r(y(x)) + C^-\;. \ \eqno(30)
$$
(30) shows, that any quantity which is $O(r(x)^{-\beta})$
($O(r(y)^{-\beta})$)\footnote{
 $f(s)=O(s^\gamma)$ is used here to denote a function satisfying
$|f(s)| \leq C(s+1)^\gamma$ for some positive constant $C$.} is
also $O(r(y)^{-\beta})$ ($O(r(x)^{-\beta})$), when composed with
$\phi_2 \circ \phi_1^{-1}$ ($\phi_1 \circ \phi_2^{-1}$). Moreover,
due to (21), (27) and (30)
$$
\partial O(r(y(x))^{-\alpha})/\partial
x^i=O(r(y(x))^{-\alpha-1})\;,\qquad
\partial O(r(x(y))^{-\alpha})/\partial y^i=O (r(x(y))^{-\alpha-1}) \ \eqno(31)
$$
((31) holding for the functions appearing in the metric). (27)
and (31) allow us to write (25) and (26) in the following form
$$
\displaystyle\sum_{k} {\partial y^k\over \partial x^i}\ {\partial
y^k\over \partial x^j} =\delta_{ij} +O(r^{-\alpha})\, \ \eqno(32)
$$
$$
\displaystyle\sum_{k} {\partial x^k\over \partial y^i}\ {\partial
x^k\over \partial y^j} =\delta_{ij} +O(r^{-\alpha})\;. \ \eqno(33)
$$
In (32)  and (33) it is irrelevant whether $O(r^{-\alpha})\
\hbox{is}\ O(r(x)^{-\alpha})\ \hbox{or}\ O(r(y)^{-\alpha})$, in
virtue of (30). Let us introduce
\begin{eqnarray*}
&&A^i{}_j = \partial y^i / \partial x^j\;,\qquad B^i{}_j =
\partial x^i/\partial y^j\;,
\\
&& C_{ijk} = A^m{}_i g^2_{m\ell} \partial A^\ell{}_j / \partial
x^k = g^2_{m\ell} (\partial y^m/\partial x^i) \partial^2 y^\ell/
\partial x^j \partial x^k\;,
\\
&& D_{ijk} = B^m{}_i g^1_{m\ell} \partial B^\ell{}_j / \partial
y^k\; .
\end{eqnarray*}
 Differentiating (26) with respect to $x$, taking into account
(27), (30) and (31) leads to
$$C_{ijk} + C_{jik} = O(r^{-\alpha-1})\;.$$
A standard cyclic permutation calculation, using the symmetry of $C_{ijk}$
in the last two indices yields
$$C_{ijk} = O(r^{-\alpha-1})\;.$$
This equality, (24), (27) and the definition of $C_{ijk}$ imply
$$
\partial^2 y^i/\partial x^j \partial x^k = O(r^{-\alpha-1})\;. \ \eqno(34)
$$
In a similar way one establishes
$$
\partial^2 x^i/\partial y^j \partial y^k = O(r^{-\alpha-1})\;. \ \eqno(35)
$$
It is elementary to show, using (32), (33), (34) and (35) that the
following quantities
$$ \mathring{A}^i{}_j = \lim_{r\to\infty} A^i{}_j (r\vec n)\;,$$
$$ \mathring{B}^i{}_j = \lim_{r\to\infty} B^i{}_j (r\vec n)\;,$$
 ($\vec n$ -any vector satisfying $\sum
(n^i)^2=1$) exist and are constant matrices ($n^i$ independent),
with $A=B^{-1}$. Define
$$
\zeta ^i(x)=y^i(x) - \mathring{A}^i{}_j\  x^j\,
$$
$$
\eta^i(y)=x^i(y) - \mathring B^i{}_j\  y^j\;. \ \eqno(37)
$$
(35) leads to
$$A^i{}_j (r_2 \vec n) - A^i{}_j (r_1 \vec
n)=\int_{r_1}^{r_2}(\partial^2 x^i (r\vec n)/\partial x^j \partial
x^k)\ n^k\ dr=O(r_1^{-\alpha})\;$$ for $r_2 > r_1$. Going with
$r_2$ to infinity, making use of (36) and (37) one obtains
$$\zeta^i{}_{,j} (x) = O(r^{-\alpha})\;,$$
which implies
$$\zeta^i (x) = O(r^{1-\alpha})\;,$$
(where $O(r^0)$ is understood as $O(\ln r))$ --- this establishes
lemma 1.

\bigskip

\parindent=0 mm
\underbar{Theorem 2}: Let $(g, \phi_a), a=1,2$, satisfy the
hypotheses of theorem 1 and lemma 1. Then

\parindent=12 mm
1) $m (g,\phi_1)= m (g,\phi_2)$

2) $P_i (g,\phi_1)= \omega_i{}^j P_j (g,\phi_2)$

$(\omega \in O(3)$, given by lemma 1).

\bigskip

\parindent=0 mm
\underbar{Proof}: Point 2) above is trivial, point 1) follows by a
well known argument from the result of lemma 1, we will repeat it
here for completeness. From lemma 1 we have
$$ k^2_{ij} = g^2_{ij} - \delta_{ij} = \mathring B^k{}_i
B^\ell{}_j k^1_{k\ell}(x(y))+ \mathring B^\ell{}_j
\eta^\ell{}_{,i}(y) + \mathring B^\ell {}_i \eta^\ell{}_{,j} (y)+
O(r^{-2\alpha})
$$
$$(k^1_{ij} = g^1_{ij}-\delta_{ij})\;. $$
Therefore
$$
\partial g_{ij}^2(y)/\partial y^j - \partial g_{jj}^2(y)/\partial y^i=\mathring
B^k{}_i (\partial k_{kj}^1(x(y)) / \partial x^j - \partial
k_{jj}^1(x(y)) / \partial x^k)$$
$$+\ (\mathring B^\ell{} _i \partial \eta^\ell / \partial y^j -
\mathring B^\ell{}_{j} \partial \eta^\ell / \partial y^i)_{,j} +
O(r^{-2\alpha-1})\;. \ \eqno(38)
$$
While integrated over the sphere $r(y)=$ const, the last term in
(38) will give no contribution in the limit $r(y) \rightarrow
\infty \ \hbox{if}\ 2\alpha + 1 > 2$, the next to last term in
(38) will give no contribution being the divergence of an
antisymmetric quantity, the first gives the ADM mass of the metric
$g_{ij}^1$ (the $B^i{}_j$ factor cancels with a similar factor
coming from the surface forms $dS_k$).

\bigskip

The condition $\alpha > 1/2$ is the best possible, in the
following sense\footnote{ This proposition is essentially due to
V.I. Denisov and V.O. Solobev [13]. Theorems 1 and 2 above show in
what sense the remaining claims of these authors are erroneous.}:

\bigskip

\parindent=0 mm
\underbar{Proposition 1}: The ADM mass of $1/2$--asymptotically
flat metrics is either infinite, or can take any value greater
than some number in the class of $1/2$--admissible coordinate
systems.

\bigskip

\underbar{Proof}: We shall establish proposition 1 for the flat
metric $ds^2 = \sum (dx^i)^2$, the general result can be obtained
by the same method. The new coordinates $y^i$ implicitly defined
by
$$x^i = (1+a \ r(y)^{-1/2}) y^i\;,\qquad a \in \R\;,$$
are easily seen to be $1/2$--admissible. The ``ADM mass" of the
flat metric in the coordinates $y^i$ can be calculated to be
$$
m = a^2/8,
$$
which establishes proposition 1.

\bigskip

Let us finally remark, that all the above results can be stated in
terms of the $H_{s,\delta}$ spaces of Y. Choquet-Bruhat and
D.~Christodoulou. The theorems of D.~Christodoulou and N.O
Murchadha [14] show that non- trivial $\alpha$--asymptotically
flat space-times satisfying Einstein equations exist with any
$\alpha
> 0$, and that the boost problem is solvable in this class of space-times
(all this holding under some supplementary conditions on the weak
derivatives of the metric). The positivity of $m $ for
$\alpha$--asymptotically flat space-times, for $\alpha > 1/2$, can
probably be established along Witten's argument lines using the
results of O. Reula [15], whose proof of existence of solutions of
Witten's equation holds in this class of metrics.

\bigskip
\parindent=0 mm

\bigskip
REFERENCES

\bigskip

[1] J. KIJOWSKI, Gen.\ Rel.\ Grav. \underbar{9}, 857 (1978).

\bigskip

[2] J. KIJOWSKI, W. TULCZYJEW, ``A symplectic framework in field
theory",Springer Lecture Notes in Physics vol.\ 107.

\bigskip

[3] M. FERRARIS, J. KIJOWSKI, Gen.\ Rel.\ Grav.\ \underbar{14},
165 (1982).

\bigskip

[4] M. FERRARIS, J. KIJOWSKI, Ren.\ Sem.\ Mat.\ Universita
Politecnico di Torino, \underbar{41}, 169 (1983).

\bigskip

[5] A. JAKUBIEC, J. KIJOWSKI, to be published.

\bigskip

[6] M. FERRARIS, J. KIJOWSKI, Gen.\ Rel.\ Grav.\ \underbar{14}, 37
(1982).

\bigskip

[7] P.T. CHRU\'SCIEL, Acta Phys.\ Pol.\  \underbar{B 15}, 35
(1984).

\bigskip

[8] P.T. CHRU\'SCIEL, Ann.\ Inst.\ H.Poincar\'e \underbar{42}, 329
(1985).

\bigskip

[9] A. SMOLSKI, Bull.\ Acad.\ Polon.\ Sci., S\'erie Sci.\ Phys.\
Astron.\ \underbar{27}, 187 (1979).

\bigskip

[10] J. KIJOWSKI, Proceedings of Journ\'ees Relativistes 1983,
Torino, eds. S. BENENTI, M. FERRARIS, M. FRANCAVIGLIA, Pitagora
Edit., Bologna 1985.

\bigskip

[11] J. KIJOWSKI, Proceedings of Journ\'ees Relativistes 1984,
Aussois, Springer Lecture Notes in Physics vol.\ 212.

\bigskip

[12] P.T. CHRU\'SCIEL, Ann.\ Inst.\ H.Poincar\'e \underbar{42},
301 (1985).

\bigskip

[13] V.I. DENISOV, V.O. SOLOBEV, Theor.\ and Math.\ Phys.\
\underbar{56}, 301 (1983).

\bigskip

[14] D. CHRISTODOULOU, N.O. MURCHADHA, Comm.\ Math.\ Phys.\
\underbar{80}, 271 (1981).

\bigskip

[15] O. REULA, Jour.\ Math.\ Phys.\ \underbar{23}, 810 (1982).
\end{document}